\documentclass[preprint2]{aastex6}

\usepackage{graphicx}
\usepackage{hyperref}
\usepackage{natbib}
\usepackage{xcolor}
\usepackage{mathtools}
\usepackage{aas_macros}
\usepackage{epstopdf}

\bibpunct{(}{)}{;}{a}{}{,}

\shorttitle{}
\shortauthors{Adhikari et al.}
\slugcomment{Accepted for publication in The Astrophysical Journal}

\begin{document}


\title{What shapes the Absorption Measure Distribution in AGN outflows}


\author{T. P. Adhikari\altaffilmark{1}, A. R\'o\.za\'nska\altaffilmark{1}, K. Hryniewicz\altaffilmark{1}, B. Czerny\altaffilmark{2}
and E. Behar\altaffilmark{3}}
\email{tek@camk.edu.pl}
\altaffiltext{1}{Nicolaus Copernicus Astronomical Center, Polish Academy of Sciences, Bartycka 18, 00-716, Warsaw, Poland}

\altaffiltext{2}{Center for Theoretical Physics, Polish Academy of Sciences, Aleja
Lotnikow 32/46, Warsaw, Poland} 

\altaffiltext{3}{Department of Physics, Technion, Haifa 32000, Israel}

\begin{abstract}
The Absorption Measure Distribution (AMD) in the X-ray outflows of Seyfert active galactic
nuclei (AGN) describes 
the distribution of absorbing column density as a function of ionization parameter. 
Up to now, AMD has been measured only for seven objects with high-resolution
X-ray data that contain absorption lines from ionized heavy elements.  
Even though the number of measured  AMDs is not large, they  display a universal
broad shape  containing a prominent dip, for which the absorbing column drops by around
two orders of magnitude. 
In this paper, we tested a range of photoionization models against the overall shape of the AMD
as observed in Seyferts. 
In particular, we demonstrate that the shape of the AMD depends both on the
spectral energy distribution (SED) of 
radiation which enters the outflow, and the density of the warm absorber (WA). 
The model that best reproduces the 
observed  shape of the AMD is  one wherein the gas density of  the WA  is of the order of 
$10^{12}$ cm$^{-3}$, irradiated by  an SED whose optical/UV luminosity  is
100  times higher than the  X-ray luminosity. When the cloud density is higher
than $\sim 10^{11}$ cm$^{-3}$, free-free heating dominates the entire absorber,
and only one instability zone occurs, which is in agreement with observations.
\end{abstract}


\section{Introduction}

The heavy element content in the intergalactic medium is constantly enriched by 
warm/hot outflows from  active galactic nuclei (AGN).
X-ray spectral studies of those objects led to the detection of many blueshifted narrow 
absorption lines from  highly-ionized elements providing a great opportunity 
to study the warm absorber (WA) in the vicinity of the super-massive black hole
(SMBH) \citep[][and other references therein]{Kaspi2001,Collinge2001,Kaastra2002,Behar2003,Netzer2003,Krongold2003,Yaqoob2003,steen2003a,Blustin2003,Rozanska2004,Turner2004,Steenbrugge2005,
Costantini2007,Winter2010,Tombesi2013,Laha2014,Laha2016,Silva2018,Mao2018}.

The observed absorption lines typically indicate ionic column densities 
of the order  of $10^{15-18}$ cm$^{-2}$ in the WA. For the given ionic column densities, the 
photoionization calculations done through thin constant-density slabs of matter allow
the derivation of equivalent hydrogen column densities taking into account appropriate values of 
the ionization parameter. Various ions indicate the distribution of 
equivalent hydrogen column densities typically in the range of $10^{18-23}$ cm$^{-2}$.
These columns correspond to the continuous change
of the ionization parameter $\xi$,
which spans  a range of few decades when determined from the data of individual source
\citep{Steenbrugge2005,Costantini2007}. 

\citet{Holczer2007} proposed to describe the ionization structure of the wind 
by determining the  absorption measure distribution (AMD), which describes how the
equivalent hydrogen column density of the 
ionizing material behaves with the change of the ionization parameter along the line of sight. 
Since then, several attempts have been made to derive the continuous ionization structure of 
the WAs in several AGN  \citep[][]{Holczer2007,Behar2009,Detmers2011} utilizing 
high-resolution X-ray spectra. The measurement of AMD strongly depends on the number of
observed absorption lines. The more observed lines,
the wider the range over which ionization parameter is determined. For each line, the equivalent width 
has to be determined precisely; this can be properly done only using 
high-energy resolution detectors such as the gratings aboard the {\it Chandra X-ray observatory} and
{\it XMM-Newton}  \citep{Steenbrugge2003,Costantini2007}. 
Furthermore, the AMD error bars are mostly obtained from  a Monte Carlo fitting method
\citep[see][for the detailed method description]{Holczer2007},
and depend only slightly on the assumed abundances in the photoionization calculations
\citep{Holczer2012}.

\citet{Holczer2007}  used {\it Chandra} observations of the WA in IRAS 13349+2438
to demonstrate that its AMD displays a
deep minimum in column density, where column density:
it decreases by 2-3 orders of magnitude,
and is consistent with negligible gas
absorption for values of $\log\xi$ between $\sim 0.8$ and 1.7 (see Eq~\ref{eq:xi_titan} for definition). 
Such deep minima are also present in the AMDs of other objects:
NGC~3783, NGC~5548, MCG$-$6-30-15, NGC~3516,
NGC~7469 \citet{Behar2009}. The overall measured normalization, defined as the average value of AMD outside
the dip, for these six sources is of the order 
of $\sim 4\times 10^{21} ~\rm cm^{-2}$. However, the case of Mrk 509 differs: \citet{Detmers2011} obtained 
two prominent dips around $\log \xi$ $\sim$ $2-3$ and $3-4$, 
with slightly lower normalization. 

To reproduce the observed  broad AMD theoretically, we consider a continuous 
ionization structure across the WA, so that  photoionization computations 
return a broad range of ionization states. Each model for the WA contains a single gas cloud under 
constant total pressure (i.e.,  the sum gas pressure and radiation pressure is a constant; hereafter CTP). 
At each column density depth, we can examine the temperature and ionization structure determined
by all relevant heating/cooling processes. In such a cloud, the  such as the
ionizing spectral energy distribution (SED), ionization parameter, and gas density are
defined only at the cloud surface. All of them change physically with cloud depth due to gas suppression by 
radiation pressure. 
This continuous ionization structure for the warm absorber under
CTP was presented by \citet{Rozanska2006}, and \citet{goncalves2007}. Later, \citet{Stern2014}
introduced  the concept of deriving the 
AMD for a  continuous ionization structure based on the assumption of a radiation pressure confinement
(RPC) of the WA material. Using the {\sc cloudy}\footnote{see \citet{Ferland2017} for the latest information on
the numerical code {\sc cloudy} } photoionization code, the latter authors were successful in
reproducing the observed normalization and the slope of the AMD for the six objects
mentioned above. 
However, they were not able to quantitatively reproduce the deep minimum of the AMD.
The difference between the RPC and CTP modes is that in the case of RPC, treated by the {\sc cloudy} code,
it is assumed that radiation pressure diminishes exponentially with
optical depth \citep[Eq.~6 in][]{Stern2014}, while in the case of CTP,  treated by the {\sc titan}
code the radiation pressure is self-consistently computed from the true intensity field
\citep{Dumont2003}. 

\citet[hereafter AD15]{Adhikari2015} demonstrated that the two dips observed in the AMD of Mrk~509
can be successfully reproduced by assuming a single WA model
under CTP using the {\sc titan} photoionization code \citep{Dumont2000}.  For the first time, AD15
confirmed that such deep minima in AMD are evidence of a thermal instability for a specific
ionization and temperature regime. However, the  observed normalization of AMD for Mrk~509
as obtained by \citet{Detmers2011} is lower by a  
factor of $\sim$ 30 than the AMD normalization 
obtained from the {\sc  titan} model of AD15 (see lower panel of Fig. 4.4 in AD15). 
\citep{Goosmann2016} performed exactly the same photoionization computations with the {\sc titan} code, to 
explain the constant-pressure WA in NGC~3783, and they achieved 
the discontinuity in AMD, but they did not continue their computations in the thermally
unstable region. 
 
In order to understand what shapes the AMD in AGN outflows, in this paper, we present a systematic study
of WAs under CTP, and focus on the importance of the SED shape and the local density at the cloud's irradiated
face in shaping the AMD.

We employ the {\sc titan} numerical code to simulate the photoionization process by computing the
thermal and ionization structure of the absorbing gas subjected to the incident radiation field of the AGN.
The resulting models, obtained for a  large grid of parameter space display thermal and ionization properties
of absorbing gas which change continuously between different gas layers.

We had previously found that the AMD normalization drops by an order of magnitude when 
the WA gas is illuminated by a SED whose optical/UV flux is higher by two orders of magnitude than 
the X-ray flux \citep{adhikaripta2018}. Here, we continue this modeling work, and find that the
overall SED shape influences the number dips preset in the AMD. in AMD. 

Knowing how SED influences the AMD normalization and the number of dips, 
we then investigate the dependence of AMD models on the density of the absorbing gas,  and find that
a high gas density, of the order of 10$^{12}$~cm$^{-3}$, 
and an ionization parameter of the order of $10^3$~erg~cm~s$^{-1}$ provides the best representation of the 
observed shape of the AMD obtained by \citet{Behar2009}. This result provides additional support that the 
outflow observed in X-rays is very dense, and it is in agreement with recent  indications that
the density of the broad line region (BLR)
is equivalently high \citep{hryniewicz2012,Adhikari2016,Adhikari2017,adhikari2018,panda2018},
suggesting that both the WA and BLR 
may originate from the upper, dense accretion disk atmospheres, even if the gas dynamics are not the same.
Interestingly, the investigation of the photoionization properties of the BLR led to estimates
of the optimal density 
of BLR being about $10^{10}$~cm$^{-3}$ \citep{Osterbrock2006}.   

The description of photoionization computations together with input parameters is presented in 
Sec.~\ref{sec:model_des}. The dependence of the AMD models 
on SED shape, ionization parameter and gas density is discussed in  
Sec.~\ref{sec:amd_dependence}, while 
Sec.~\ref{subsec:gen_amd_comp} contains the 
comparison of our model with the observed AMD shape. Finally, the discussion and conclusions 
are presented in the Sec.~\ref{sec:discussion} and \ref{sec:conclusion} respectively.

\section{Photoionization models}
\label{sec:model_des}

We computed the photoionization models using the 
numerical code {\sc titan} \citep{Dumont2000,Rozanska2004} 
under the assumption that a X-ray absorber in AGN is 
in total pressure equilibrium, sustained by the CTP condition.
As demonstrated in earlier works \citep{Dumont2003,Rozanska2006,Adhikari2015}, {\sc titan} uses 
the ALI \citep[accelerated lambda iteration;][]{paletou1995}  method of radiative transfer computations and
allows us to well resolve the thin layers of strong 
temperature and density gradients that develop in the photoionized gas cloud.

In this paper, we consider a CTP single gas cloud defined by the following parameters:  
number density $n_{\rm H,0}$ and 
ionization parameter $\xi_{0}$
both assumed at the illuminated surface, and the total column density
$N_{\rm H}^{\rm tot}$ \citep[see Sec. 3 of][for 
the general description of parameters in {\sc titan} 
photoionization computations]{Adhikari2015}.  
Due to  illumination by the AGN radiation field from one side, such a cloud forms a highly stratified medium
with strong gradients in density, temperature and ionization state.
The cloud stratification is calculated self-consistently by solving the ionization and
thermal balance across the cloud. 

The ionization parameter strongly varies with depth through the cloud, and we calculate its value within a
cloud as a function of the optical depth measured from the illuminated cloud using the expression
\begin{equation}
\label{eq:xi_titan}
\xi = 4 \, \pi c\, k\, T \, {P_{\rm rad} \over P_{\rm gas}} = 4 \, \pi c\, {P_{\rm rad} \over n_{\rm H}},
\end{equation}
where $c$ is the velocity of light, and $k$ is Boltzmann's constant. In the above definition, 
the ionized radiation is integrated over the whole range of the ionizing continuum as described below. 
This differs from the original definition, where ionized radiation is integrated over 
the hydrogen ionizing continuum from 1-1000~Ry \citep{Krolik1981}. Such a difference is always taken 
into account in this paper when introducing the value of the input ionization parameter. 

The temperature $T$, number density $n_{\rm H}$, radiation pressure $P_{\rm rad}$ and gas
pressure $P_{\rm gas}$ are
computed self-consistently by {\sc titan} at each layer of the photoionized gas cloud.
The values of the local density and gas pressure take into account all partial ionization states for the fractional 
abundances obtained from the ionization and thermal balance. 
The radiation pressure is self consistently derived from the second moment of the radiation field after checking that radiation balance is sustained globally. 

For each computed model, the 
AMD is derived taking into account the stratification of the ionization parameter $\xi$,
using the following relation:
\begin{equation}
\label{eq:amd}
  {\rm AMD}= \frac{dN_{\rm H}}{d(\log\xi)}=2.303~\xi\frac{dN_{\rm H}}{d \xi}.
\end{equation}
We compute a large grid of models for initial gas parameters defined at the cloud surface:
$\xi_{0}$ from $10^2$ up to $10^5$~erg\,cm\,s$^{-1}$ with logarithmic grid step equal to 1, 
and $n_{\rm H,0}$ from $10^8$ up to $10^{12}$~cm$^{-3}$ with the same grid step. 
The total column density of the clouds $N_{\rm H}^{\rm tot}$ spans from $10^{20}$ to $10^{23.5}$~cm$^{-2}$ 
with logarithmic grid step equal to 0.5. The most interesting clouds are those of large total column densities, 
since they include all ionization fronts, and their temperature structure goes from high Compton 
temperature ($\sim 10^7$~K) down to the lowest value ($\sim 10^4$~K) on the non-illuminated side of the cloud. 
In particular, it means that the gas hosts a broad range in ionization degree. 

In order to investigate what shapes the AMD in AGN,
we employed various types of AGN SEDs (Section \ref{sec:sed_shapes})
in our computations together with the above  values of ionization parameter, the gas density on the cloud surface, 
and the total gas column density, as discussed above.
Since our grid of models is already large -- up to 800 cases -- we did not explore the effect of metallicity here, 
fixing it at the solar value. 
Selected important results of our photoionization computations are presented in 
Sec.~\ref{sec:amd_dependence}. 

\subsection{General SED shapes}
\label{sec:sed_shapes}
An  important input parameter in photoionization simulations is the shape of 
the ionizing SED, which influences the 
 thermal and ionization structure of the WA cloud.
Studies of the  stability curve for different shapes of the 
radiation field have revealed that the resulting curves are significantly 
different, and in particular very sensitive to the UV/soft X-ray part of the 
SED \citep{Rozanska2006,Rozanska2008,Chakravorty2009,Chakravorty2012}.
If the ionizing continuum is 
sufficiently soft, i.e. dominated by the UV/soft X-ray band, the thermal and ionization states of 
WA are sensitive to their density apart from the usual dependence on the ionization parameter
 \citep{Rozanska2008,Chakravorty2009}.

\begin{figure}[]
\centering
\hspace{-0.6cm}
\includegraphics[width=9.0cm]{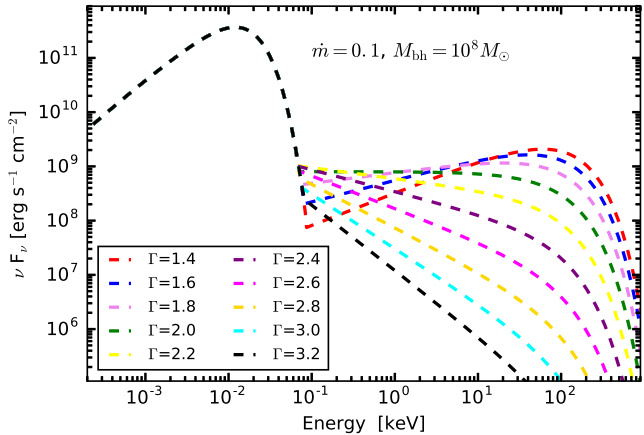}
\caption{SED A shapes dominated by the big blue bump (optical/UV) component.
Cases with different X-ray photon indexes are shown. The disk luminosity $L_{\rm disk}$ is always 
100 times higher than the X-ray luminosity $L_{\rm X}$.}
\label{fig:sed_ls_hm_all_g}
\end{figure}

To make systematic studies of the AMD, we consider a range of SED shapes, 
assuming that the overall spectrum originates from the AGN nucleus.
This emission is represented by 
two major components that contribute 
to the broad AGN spectrum: a multi- black body disk spectrum and an X-ray
power law with an exponential cut-off.
The variation in SED shapes is achieved by varying the 
dimensionless mass accretion rate  relative to the Eddington rate $\dot{m}$, and the power-law photon index 
 $\Gamma$. The mass of the SMBH is taken as $M_{\rm BH}$ = $10^{8}$ M$_{\odot}$, and the high energy cut-off 
 $E_{\rm cut}=100$~keV in all cases. 
 
\begin{figure}[]
\centering
\hspace{-0.6cm}
\includegraphics[width=9.0cm]{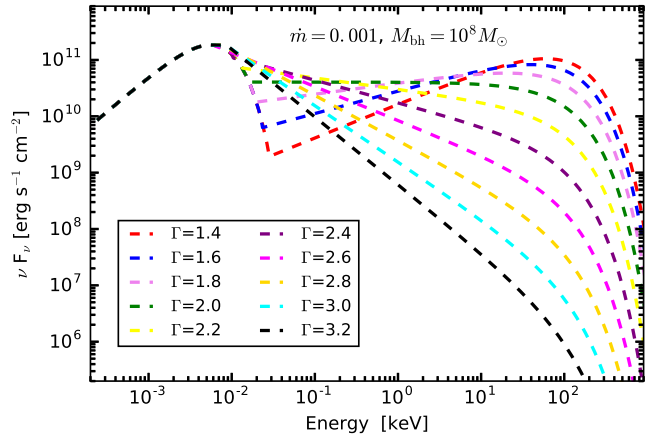}
\caption{SED B shapes, where the disk emission $L_{\rm disk}$ is comparable to X-ray emission
$L_{\rm X}$.Cases with different the X-ray photon indexes are shown.}
\label{fig:sed_hs_lm_all_g}
\end{figure}

Since the incident radiation has already two free parameters, $ \dot m$ and $\Gamma$, and considering 
the  input parameters for the warm absorber ($\xi_0$, $n_{\rm H,0}$, $N_{\rm H}^{\rm tot}$),
the number  of computed models is very large, and each model requires 24 hours CPU time.
So, for clarity, 
we present here only those models that we found to be most crucial in shaping the AMD. In particular,  
we divided our models into two groups of SED which differ by the strengths of the 
black body disk emission and by the relative normalization of disk to X-ray power-law  
emission. The first group, named SED~A (depicted in Fig.~\ref{fig:sed_ls_hm_all_g}), represents the case dominated by 
the big blue bump component, and it is calculated for $\dot{m} = 0.1$.
The second group, named SED~B (depicted in Fig.~\ref{fig:sed_hs_lm_all_g}), has black body disk emission 100 times weaker 
than for SED~A, and it is calculated for $\dot{m} = 0.001$. 

In the case of both groups of models A and B, 
the variation in the X-ray power-law slope is done by taking 
ten values of photon index $\Gamma$ from 1.4 up to 3.2 with a grid step of 0.2.
The observed values of $\Gamma$ for typical AGN are between 1.5--2.5 \citep[][and references therein]{feng2014}.
  Our range is slightly wider to emphasize the importance of hard X-ray photons on the high-ionization end of AMD. 
While varying the X-ray photon index, the X-ray luminosity, resulting from integration
of the power-law shape from 0.001 to 100~keV, is kept constant
for a given value of $\dot{m}$. This condition is achieved by 
varying  $\alpha_{\rm OX}$ parameter \citep{Tananbaum1979}. $\alpha_{\rm OX}$  describes the ratio of flux 
emitted at {$2500$~\AA} to the flux emitted in soft X-rays at $2$~keV, and it is 
defined by the relation
\begin{equation}
\alpha_{\rm OX}=-0.384 \log \left(\frac{F_{\rm 2 keV}}{F_{\rm 2500\AA}} \right).
\label{eq:alpha_ox}
\end{equation}
For group A SEDs, the black body disk component is always two orders of magnitude 
 more luminous than the X-ray power law, while for group B SEDs, 
luminosities of both components are equal.
The total luminosity depends mostly on the disk accretion rate. 
The resulting two groups of parameters and corresponding luminosities 
are presented in Table~\ref{tab:param_gen_SEDs}. 

\begin{table}[!h]
\caption{\small The parameters used for computing the general SEDs.
The mass accretion rate relative to Eddington $\dot{m}$, disk luminosity $L_{\rm disk}$
in erg~s$^{-1}$, the X-ray luminosity $L_{\rm X}$ in erg~s$^{-1}$, the 
photon index $\Gamma$, and optical --X-ray index $\alpha_{\rm OX}$
are presented in columns 2, 3, 4, 5 and 6 respectively. The mass of the 
SMBH in all cases is taken as $M_{\rm BH}$ = $10^{8}$ M$_{\odot}$.}
\begin{center}
\label{tab:param_gen_SEDs}
\begin{tabular}{cccccc}
\hline 
SED& $\dot{m}$&$L_{\rm disk}$& $L_{\rm X}$& $\Gamma$ & $\alpha_{\rm OX}$\\
 & & $\times 10^{43}$ & $\times 10^{43}$ & &  \\
\hline
A  & 0.1 & 126 & 1.26 & 1.4 -- 3.2  & -1.9 -- -2.8 \\
B  & 0.001 & 1.26 & 1.26 & 1.4 -- 3.2    &-1.3 -- -2.1\\ 
\hline 
\end{tabular}
\end{center}
\end{table}
       

\section{Initial parameters and AMD shape}
\label{sec:amd_dependence}
In order to demonstrate most clearly how different SEDs affect the resulting AMD models,
 we fixed parameters at the each cloud surface to be $n_{\rm H,0}=10^{8}$~cm$^{-3}$ and
$\xi_{0}=1.36\times10^4$~erg~cm~s$^{-1}$. For both groups A and B,  only models with a broad ionization distribution 
are shown for various values of X-ray power law photon index 
$\Gamma$. The models with low values of initial total column density and with extreme values of $\Gamma$
exhibit narrow AMDs.
 
The temperature structure displays a very steep gradient at those values of the local gas column density $N_{\rm H}$  where 
ionization instability occurs in the gas under constant pressure \citep{Rozanska2008,Adhikari2015}. 
Below we describe the cloud structure in detail, separately for these two different SEDs.
Then, for the given SED, we explore how the cloud structure depends on the  initial ionization parameter and the gas density.

\subsection{SED A}
\label{sec:sed_a}

\begin{figure}[!h]
\centering
\hspace{-0.4cm}
\includegraphics[width=8.9cm]{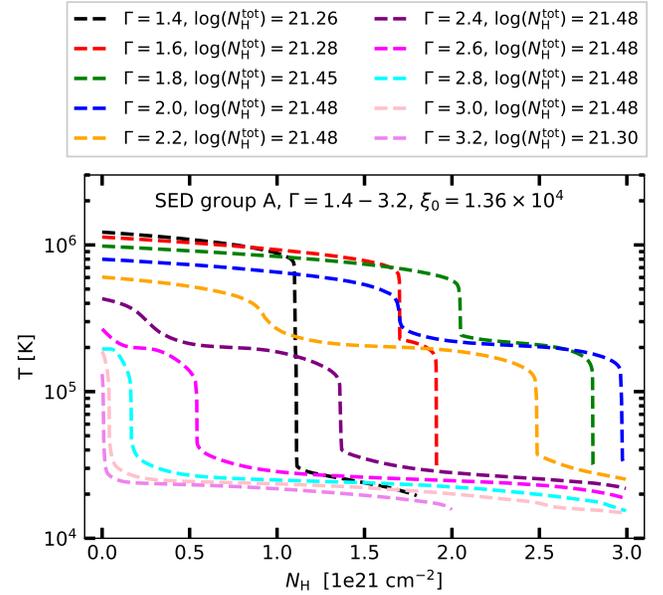}
\caption{Thermal structure across a clouds illuminated with  SED A,
for different values of $\Gamma$ in the range 1.4--3.2. 
The models illustrate the temperature behavior along the line of sight through the 
entire cloud, from the illuminated face ($N_{\rm H}^{\rm tot}=0$; highest temperatures) 
to the non-illuminated face (maximum $N_{\rm H}^{\rm tot}$; lowest temperatures). 
The values of 
$n_{\rm H,0}=10^{8}$~cm$^{-3}$ and $\xi_{0}=1.36 \times 10^{4}$~erg~cm~s$^{-1}$ at the cloud surface are fixed
in these computations.}
\label{fig:structure-ls-hm-n8-xi5_allg}
\end{figure}

The temperature variations across
the depths of the clouds for group A SEDs are shown 
in Fig.~\ref{fig:structure-ls-hm-n8-xi5_allg}. The local column density $N_{\rm H}$
is measured from the illuminated surface of the cloud. 
Steep drops at specific values of the column density are well visible.
The number of steep temperature drops is one or two depending on the spectral slope of the incident hard X-ray component. 
The value of surface temperature, which is a result of thermal balance,  differs by one
 order of magnitude,  from $1.58 \times 10^{6}$ to $1.58 \times 10^{5}$~K, between the 
most extreme values of $\Gamma$ values i.e. 1.4 and 3.2. 
The corresponding values  of the Compton temperature (mean photon energy of 
illuminating continuum) for those models is
 $1.75 \times 10^6$ and $3.27 \times 10^4$~K respectively. 
As the  value of 
$\Gamma$ increases, the number of sudden temperature drop
changes from two to one. 

For $\Gamma \leq 2.4$, there are two distinct 
regions where a quick drop in the temperature happens. 
However for  $\Gamma > 2.4$, only a single temperature drop is seen since the temperature at the surface of the cloud is too low to form a well developed hot region. However, this division is very dependent on the SED shape and cannot be treated as a general behavior. 

\begin{figure}[!h]
\includegraphics[width=8.8cm]{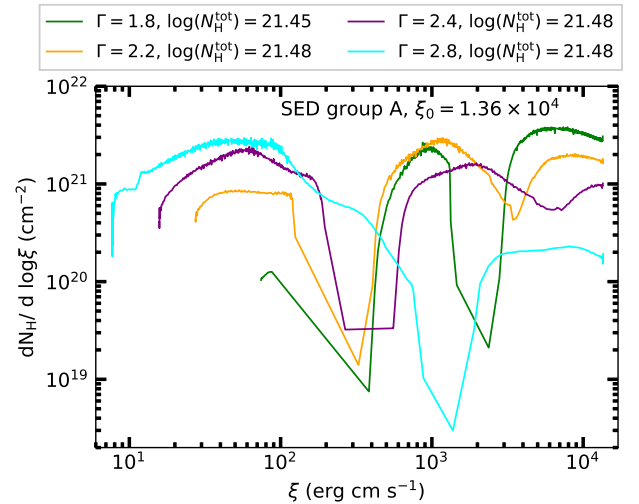}
\caption{AMD models for  SED group A for various values of $\Gamma$. 
The AMDs are computed for all the models with various values of 
$N_{\rm H}^{\rm tot}$ until the broad ionization distribution is obtained. 
Values of $n_{\rm H,0}=10^{8}$~cm$^{-3}$ and 
$\xi_{0}=1.36 \times 10^{4}$~erg~cm~s$^{-1}$ 
are used in the model computations. A single AMD dip is present only for $\Gamma > 2.4$.}
\label{fig:amd_SEDA_many_xi5_selg}
\end{figure}

The sudden drops in the temperature are also
nicely reflected by the switch in the number
of prominent AMD dips, from two to one, in the 
corresponding shapes of AMDs in the  group A models as shown in 
Fig.~\ref{fig:amd_SEDA_many_xi5_selg} for selected values of $\Gamma$. The models clearly demonstrate that the number of AMD dips depends on the slope of the illuminating
X-ray power law. This effect is directly reflected in the value of temperature at the cloud surface, which is derived from the energy balance equation. When the X-ray illuminated
continuum is relatively harder (lower $\Gamma$), more energetic  photons heat gas 
more strongly in photoionization and Compton or free-free processes.  

The common feature seen in all AMDs  for  group A models, despite variations in 
$\Gamma$, is that the overall normalization always remains at the level of 
$\sim$~few~$\times 10^{21}$~cm$^{-2}$. This result is in very good agreement with the 
AMD normalization for Seyfert galaxies  obtained from observations \citep{Behar2009,Detmers2011}. It also demonstrates that the AMD normalization level is directly connected to the total column density of the gas stream displaying thermal instability while remaining  under constant pressure \citep{Rozanska2006}.
Nevertheless, the positions of drops caused by thermal instability do not agree with 
those in the observed AMDs.

\begin{figure}[!h]
\centering
\includegraphics[width=8.8cm]{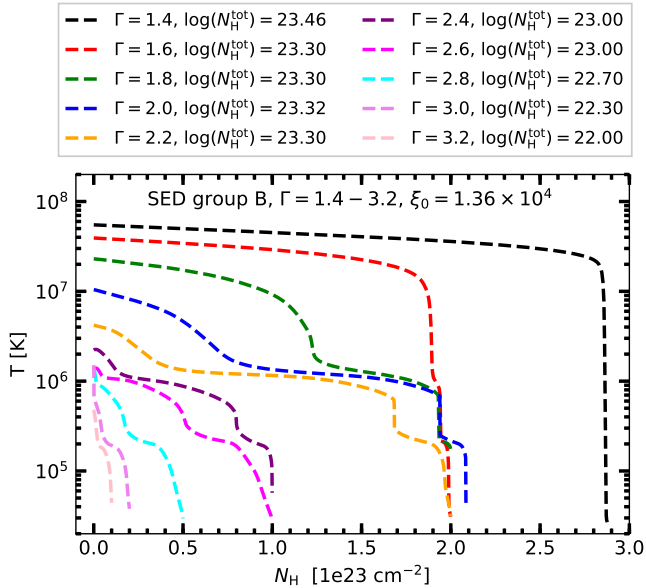}
\caption{Thermal structure across the clouds
illuminated with the SED B, for different values of $\Gamma$ in the range 1.4--3.2.
Presented models are computed for 
$N_{\rm H}^{\rm tot}$ large enough to reach the lowest temperatures on the non-illuminated side of the cloud. The values of 
$n_{\rm H,0}=10^{8}$~cm$^{-3}$ and $\xi_{0}=1.36 \times 10^{4}$~erg~cm~s$^{-1}$ at the cloud surface are fixed
in these computations.}
\label{fig:structure-hs-lm-n8-xi5-gall}
\end{figure}


\subsection{SED B}
\label{sec:sed_b}

The temperature structure in the case of SED B models as a function of photon index in 
the range $\Gamma = 1.4-3.2$ is shown 
in Fig.~\ref{fig:structure-hs-lm-n8-xi5-gall}. 
From this figure, a clear separation between 
two types of models is seen. For values of  $\Gamma$ in the range 1.4--2.2, 
a large amount of absorbing material $\geq 2\times10^{23}$ cm$^{-2}$  is required in order
to span  the broad range of ionization levels and reach the minimum temperature.
However, for value of $\Gamma >$ 2.2, the total column density required 
is $\leq 10^{23}$~cm$^{-2}$. These clouds  also have much a lower  
value of Compton temperature at the cloud surface. The value of the Compton 
temperature for the hardest spectrum,  
$\Gamma=1.4$, is $8.7 \times 10^7$~K, while for the softest spectrum, 
$\Gamma=3.2$, the Compton temperature is $4.7 \times 10^4$~K.

We point out here that for the same value of ionization parameter, SED A's illuminating continuum heats up the cloud  to a temperature an order of magnitude
lower compared to  SED B. This fact clearly indicates that the ionization parameter is not the direct indicator of the gas temperature, the shape of the SED of the ionizing photons that interact with gas on the microscopic level has a major influence on the final thermal equilibrium. Furthermore, the large number of X-ray photons in SED B causes over-ionization of  the gas, decreasing the number of ions that can produce absorption. Therefore, to achieve the observed quantity of absorbing ions, we 
need to increase total column density. For this reason, the SED B models always require
 high values of total column densities,  $\sim 10^{23}$~cm$^{-2}$, more than an order of magnitude higher than the column densities indicated by the observed AMD shape \citep{Behar2009,Detmers2011}. This effect was demonstrated in a recent work by \citet{adhikaripta2018}.  

\subsection{Ionization parameter}
\label{sec:ionpar}

 \citep{Adhikari2015} successfully fitted the positions of two observed AMD drops 
 with the CPT model, though just for case of Mrk~509.
Since the aim of the current  paper is to explain the  the shape of the AMD in  six sources as observed by \citet{Behar2009}, we assume group A SEDs hereafter. 
There may, of course, exist an intermediate SED shape (between A and B) 
which will yield the optimum match with the observed AMD. 
To  extract that optimal SED  shape, we would need to  calculate a large number of models and use machine learning to compare models with observations, and we defer that process to a future work.

\begin{figure}[!h]
\centering
\includegraphics[width=8.7cm]{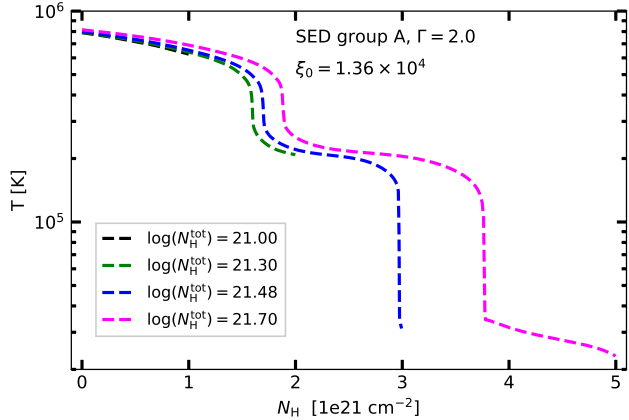}
\caption{Thermal structure across the cloud  
illuminated with SED A, $\Gamma$=2.0.
Models are computed for $n_{\rm H,0}=10^{8}$~cm$^{-3}$ and
various values of $N_{\rm H}^{\rm tot}$ as marked in the figure. 
The ionization parameter is always 
$\xi_{\rm 0}=1.36\times 10^{\rm 4}$~erg~cm~s$^{-1}$.}
\label{fig:structure-ls-hm-n8-xi5-g20}
\end{figure}

\begin{figure}[!h]
\centering
\includegraphics[width=8.7cm]{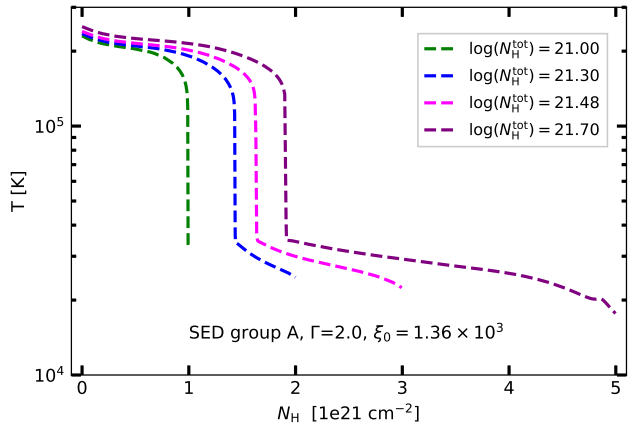}
\caption{Same as in Fig.~\ref{fig:structure-ls-hm-n8-xi5-g20} but for a surface 
ionization $\xi_{\rm 0}=1.36\times 10^{\rm 3}$~erg~cm~s$^{-1}$.}
\label{fig:structure-ls-hm-n8-xi4-g20}
\end{figure}

In this subsection, we present the influence of the values of ionization parameter and 
gas density assumed at the cloud surface on the overall AMD shape for group A SEDs. For that purpose we compared two values of $\xi_{\rm 0}$ an order of magnitude apart, $1.36\times 10^{\rm 4}$~erg~cm~s$^{-1}$ and $1.36\times 10^{\rm 3}$~erg~cm~s$^{-1}$.
In the case of the high value of $\xi_{\rm 0}$, relatively deeper gas layers are hot, of the order of Compton temperature. In this case, when ionization parameter  decreases with depth, the temperature  structure displays two steep drops
as seen in Fig.~\ref{fig:structure-ls-hm-n8-xi5-g20}. In these models, we adopted 
$\Gamma = 2.0$, $n_{\rm H,0}=10^8$~cm$^{-3}$, 
and $\xi_{\rm 0}= 1.36\times 10^{\rm 4}$~erg~cm~s$^{-1}$ only when the total column density of the cloud is high enough that a deeper cold zone develops. 
In contrast, for the lower value of $\xi_{\rm 0}$,
the deeper layers are not so hot, and the  overall 
temperature structure displays at most one steep drop, as shown
in Fig.~\ref{fig:structure-ls-hm-n8-xi4-g20}, for the same SED and gas density. 
 
\begin{figure}[!h]
\centering
\includegraphics[width=8.7cm]{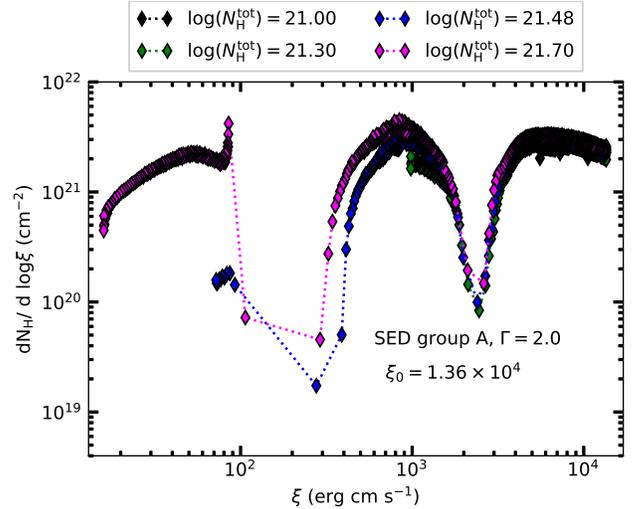}
\caption{AMDs for the same model parameters as in 
Fig.~\ref{fig:structure-ls-hm-n8-xi5-g20}.}
\label{fig:amd_SEDA_many_g20_xi5}
\end{figure}

\begin{figure}[!h]
\centering
\includegraphics[width=8.7cm]{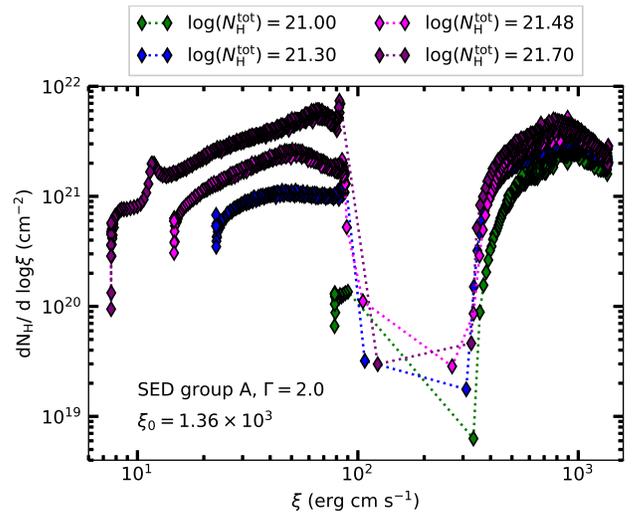}
\caption{AMDs for the same model parameters as in
Fig.~\ref{fig:structure-ls-hm-n8-xi4-g20}.}
\label{fig:amd_SEDA_many_g20_xi4}
\end{figure}

The number of drops in each case is nicely reflected in the shapes of the AMD,
 shown for the same two cases in 
Figs.~\ref{fig:amd_SEDA_many_g20_xi5} and \ref{fig:amd_SEDA_many_g20_xi4} respectively.
When we start our computations with the higher value of $\xi_{0}$, two 
dips in the AMD are clearly seen, 
  but the observations of the six sources all do not  cover AMD at
  log$\xi > 3.3$ (see Sec.~\ref{subsec:gen_amd_comp}). Therefore, to reproduce the data, it is useful to start the model with the lower value of $\xi_{\rm 0}$.
This lower ionization parameter 
potentially provides  the one prominent dip observed in the six AMDs.  
However, the dip observed in AMDs is located at  
$0.8 < \log \xi< 1.7$ which is in strong disagreement with 
our model: the model AMD, shown in Fig.~\ref{fig:amd_SEDA_many_g20_xi4}, possesses one dip at $2 < \log \xi < 2.5$. 
 
\begin{figure}[!h]
\centering
\includegraphics[width=8.7cm]{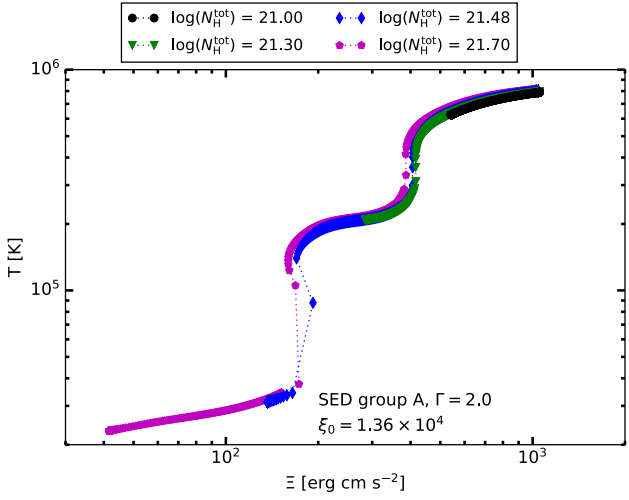}
\caption{Stability curves: $T$ versus $\Xi$ given by Eq.~\ref{eq:Xi}, for the same model parameters as in 
Fig.~\ref{fig:structure-ls-hm-n8-xi5-g20}.}
\label{fig:scurve_xhi}
\end{figure}

\begin{figure}[!h]
\centering
\includegraphics[width=8.7cm]{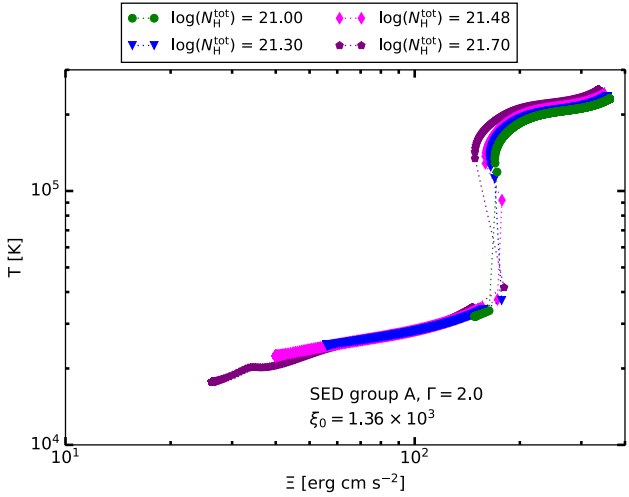}
\caption{Stability curves: $T$ versus $\Xi$ given by Eq.~\ref{eq:Xi}, for the same model parameters as in 
Fig.~\ref{fig:structure-ls-hm-n8-xi4-g20}.}
\label{fig:scurve_xlo}
\end{figure}

To confirm that thermal instabilities are responsible for the appearance of the dip  in AMD,
 we plot the stability curve, i.e., temperature versus dynamical ionization parameter $\Xi$ \citep{Krolik1981}, which relates to the parameter $\xi$ by the following equation:
\begin{equation}
\Xi = \frac{P_{\rm rad}}{P_{\rm gas}} =   \frac{\xi}{4 \pi c k T }. 
\label{eq:Xi}
\end{equation}
For the AMD presented in Figs.~\ref{fig:amd_SEDA_many_g20_xi5} and \ref{fig:amd_SEDA_many_g20_xi4} we plotted the corresponding stability curves in Figs~\ref{fig:scurve_xhi} and \ref{fig:scurve_xlo} respectively.
The dips in the AMD  appear exactly in places where unstable branches of the stability curve develop.

\begin{figure}[!h]
\centering
\includegraphics[width=8.0cm]{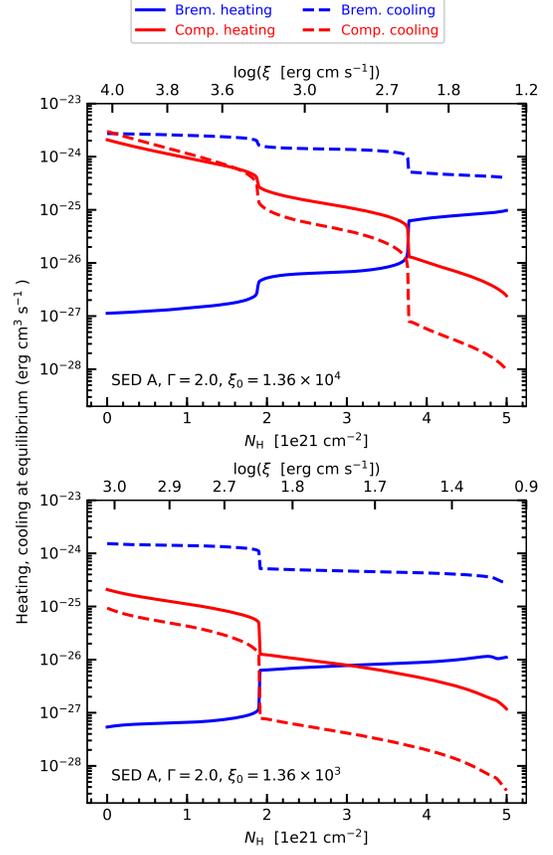}
\caption{The free-free (blue) and Compton (red) contribution to the heating (solid line) 
and cooling (dashed line)
 of the irradiated gas versus column density. All models are computed for the same SED and with $n_{\rm H,0}=10^{8}$ cm$^{-3}$ and $N_{\rm H}^{\rm tot}=5 \times 10^{21}$ cm$^{-2}$. The upper panel represents an ionization parameter one  order of magnitude higher than in the  lower panel.}
 \label{fig:heating_cooling_SED_A_den8}
 \end{figure}

To explore the dip's origin, we compare the  major gas heating-cooling processes i.e., 
Compton and  free-free, for two models of different ionization state, as 
shown in Fig.~\ref{fig:heating_cooling_SED_A_den8}. It is clear 
that when the ionization degree is higher i.e., 
$\xi_{\rm 0}= 1.36\times 10^{\rm 4}$~erg~cm~s$^{-1}$ (top panel), the Compton
contribution to the cooling (red dashed line) is comparable to the 
free-free cooling (blue dashed line) at the surface of the gas cloud. 
The situation differs as we
move to the lower ionization parameter, $\xi_{\rm 0}= 1.36\times 10^{\rm 3}$~erg~cm~s$^{-1}$ (bottom panel), where the Compton cooling contribution decreases by an order
of magnitude. At the values of the ionization parameter corresponding to the 
positions of the AMD dips, the sudden
changes in the heating-cooling rates are also observed,  
as demonstrated in Fig.~\ref{fig:heating_cooling_SED_A_den8}.
These positions correspond to the sudden temperature drops 
shown in the thermal structures in Figs~ \ref{fig:structure-ls-hm-n8-xi5-g20} 
and \ref{fig:structure-ls-hm-n8-xi4-g20}, respectively.

In the  case of the lower value of $\xi_{\rm 0}$  (bottom panel 
of Fig.~\ref{fig:heating_cooling_SED_A_den8}), there is only one sudden
change present in the heating-cooling vertical cloud structure.
 These results are again the confirmation that 
the AMD minima are caused by thermally unstable regions 
where an abrupt change in the heating-cooling processes also happens. These unstable regions are located at the same range of ionization parameter  inside the cloud (as indicated by the upper labels on both panels), independently from the value of this parameter at the front of the cloud.
Therefore, the the value of the ionization parameter at the  surface 
changes the number of dips in the AMD, but does not influence the position of those dips.

\subsection{Gas density}
\label{sec:density_effects}

The last parameter which may influence the AMD dip position and depth is 
the gas density  at the irradiated  face of the cloud, $n_{\rm H,0}$. 
To investigate this influence, all models considered here are computed with  SED A,
$\Gamma=2.0$, and the same low ionization parameter $\xi_{0}$
 ($1.38 \times 10^{3}$ erg cm s$^{-1}$), that yields one prominent dip in AMD model. 
The only parameter which we vary now is the value of the density
at the illuminated face of the cloud. 
The calculations are done 
for values of $n_{\rm H,0}=10^{8}, 10^{10}, 10^{11}$ and 
$10^{12}$ cm$^{-3}$.

\begin{figure}[!h]
\centering
\includegraphics[width=8.7cm]{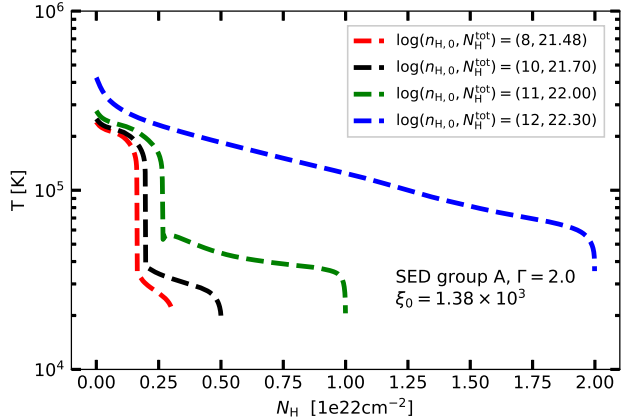}
\caption{Thermal structure across a cloud 
illuminated with  SED A, and $\Gamma$=2.0.
Four examples of different values of density 
at the illuminated face of the cloud $n_{\rm H,0}$ are shown. Densities and the 
corresponding values of total cloud column densities are listed in the figure.}
\label{fig:structure-sedA-nvar-xi4-g20}
\end{figure}

\begin{figure}[!h]
\centering
\includegraphics[width=8.9cm]{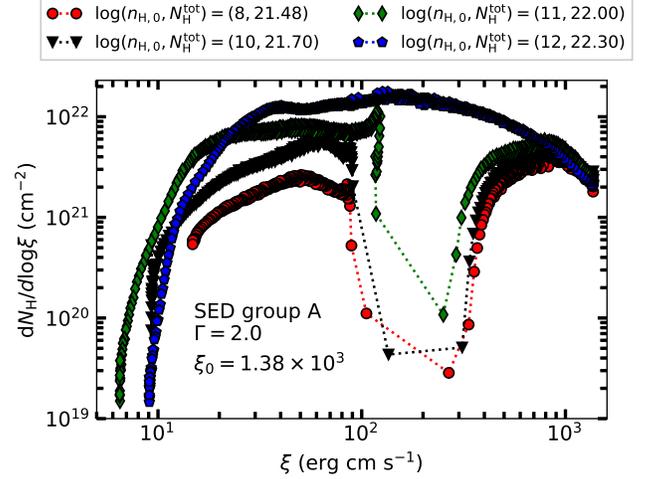}
\caption{AMDs for the same model parameters as in
Fig.~\ref{fig:structure-sedA-nvar-xi4-g20}.}
\label{fig:amd_lshm_SEDA_g20_xi4_nvar}
\end{figure}

\begin{figure}[!h]
\centering
\includegraphics[width=8.7cm]{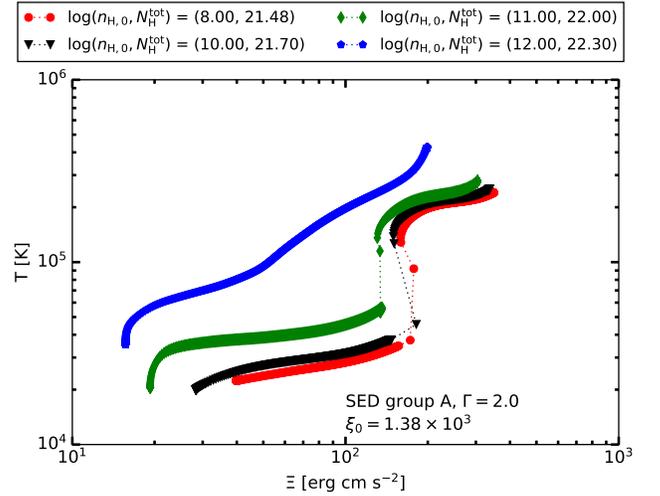}
\caption{Stability curves: $T$ versus $\Xi$ given by Eq.~\ref{eq:Xi}, for the same model parameters as in 
Fig.~\ref{fig:structure-sedA-nvar-xi4-g20}.}
\label{fig:scurve_nvar}
\end{figure}

The cloud temperature structures, the corresponding AMD models, and stability curves  are presented in Figs.~\ref{fig:structure-sedA-nvar-xi4-g20},
\ref{fig:amd_lshm_SEDA_g20_xi4_nvar}, and \ref{fig:scurve_nvar} respectively.
There is no significant difference in the overall
properties of the AMD for $n_{\rm H,0}$ $\leq 10^{11}$~cm$^{-3}$.
For these models, only
one AMD dip is present since the initial value of ionization parameter 
is already below the value $\xi \sim 3 \times 10^3$~erg~cm~s$^{-1}$, at which the first dip (counting from the cloud illuminated face) occurs in the AMD models.
However, the depth of the AMD dips as well as the AMD normalization are 
slightly different even for these range 
of densities. With increased gas density, the depth of the AMD dip 
decreases and the overall normalization increases as seen in 
Fig.~\ref{fig:amd_lshm_SEDA_g20_xi4_nvar}. As the gas density becomes large ($10^{12}$ cm$^{-3}$),  the AMD dip at $\xi \sim 200 $ disappears and a continuous AMD is obtained 
for such a model.

However, in all models we see a very strong decrease in the  AMD below $\xi \sim 10$. This decrease is not followed by a recovery at further lower values of the ionization parameter 
since our computations for assumed initial parameters
 stop when gas reaches the very low temperature limit in the {\sc titan} code, which is 8000 K. The examples we computed  for the case of high gas density, i.e.  $n_{\rm H,0} \ge 10^{10}$ cm$^{-3}$,
 are already thick enough such that the temperature
at the non-illuminated cloud face drops suddenly. 
From the temperature structure one may expect that the strong temperature 
drop that  causes the strong decrease of the AMD below $\xi \sim 10$ is also caused by thermal instability.  With the {\sc titan} code, it is not possible to go deeper into the illuminated thick gas while allowing for still a higher total column density,
since the required radiative and hydrostatic equilibrium is not achieved for thicker clouds. Therefore, in the case of the dense clouds 
presented in Fig.~\ref{fig:structure-sedA-nvar-xi4-g20}, we are not able to reconstruct 
the stable branch of the cold gas, $\sim 10^4$~K. Additional processes (probably many more line transitions) should be included in the radiative transfer calculations to solve this 
problem, and we leave  this issue to be addressed in a future work.

\begin{figure*}
\centering
\includegraphics[width=18.5cm]{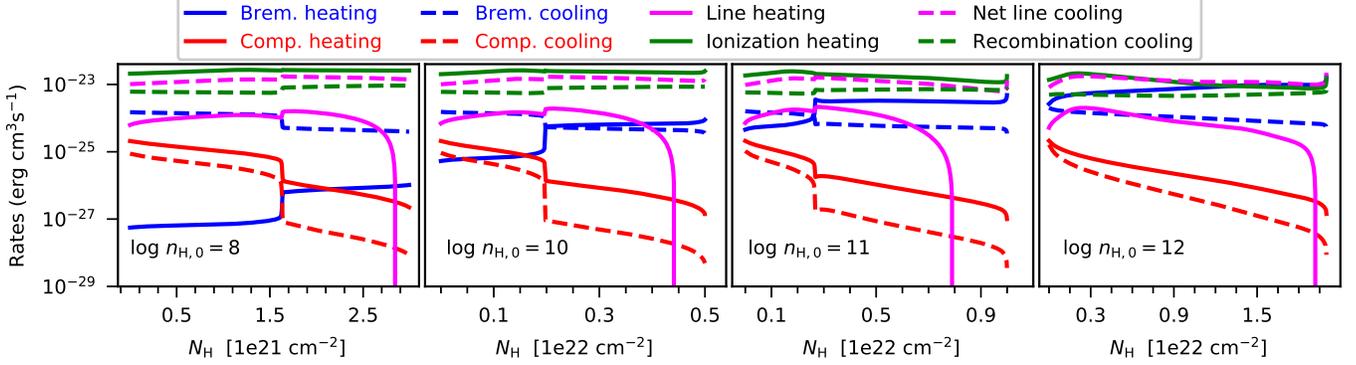}
\caption{Heating and cooling processes operating in the
gas clouds described by the parameters given in  
Fig.~\ref{fig:structure-sedA-nvar-xi4-g20}. The panels from left to right 
represent gas densities from 10$^{8}$ cm$^{-3}$  to 10$^{12}$ cm$^{-3}$.
Bound-free (ionization and recombination) and bound-bound (lines) processes stay constant over the cloud structure.
The free-free heating-cooling processes become more significant in    
relatively higher-density gas.}
\label{fig:heating_cooling_SED_A_var_den}
\end{figure*}

The importance of Compton, free-free, bound-free (ionization and recombination) 
and bound-bound (lines) heating/cooling rates  for each model  with illumination  by SED A and with different 
gas densities is shown in  Fig.~\ref{fig:heating_cooling_SED_A_var_den}.
The figure demonstrates that the physical reason for this AMD behavior is reflected
in the domination of cooling and heating processes across the cloud. 
For gas at lower density, $10^8$~cm$^{-3}$ (left panel), there is a clear domination 
of Compton heating over free-free heating. When the density at the cloud  
surface increases, the dominance of free-free heating 
over the Compton heating is clearly visible
inside the cloud for $n_{\rm H,0}=10^{10}$~cm$^{-3}$ (second  panel from the left),
and even at the surface of the cloud for densities $10^{11}$ and $10^{12}$~cm$^{-3}$ (two right panels).
For $n_{\rm H,0} \ge 10^{11}$~cm$^{-3}$ the free-free domination 
occurs across the whole absorber.
The sum of ionization and line heating is roughly linear across clouds.
Only the line heating rate decreases strongly on the non-illuminated face of the cloud,
but it is 1.5 orders of magnitude lower than the ionization heating rate.  

All dips in the AMD are related to the rapid change in the heating/cooling mechanism with 
the density, as shown here for the $\xi \sim 200$ feature, but they are less easy to demonstrate. The dip at higher $\xi$
which appears in Fig.~\ref{fig:amd_SEDA_many_g20_xi5} is related to the increased emission line
cooling in comparison with Compton cooling, and the dip at $\xi \sim 10$ is related to the onset
of heating mostly through numerous lines from atoms and (perhaps) molecules. 

The change of the dominant heating mechanism 
from Compton heating to free-free heating with a decrease in
 the density may be connected
with wind launching mechanisms. When density is relatively low, we deal with 
Compton-heated winds, but for higher density we may encounter  thermally-driven winds. 
Nevertheless, this hypothesis needs further considerations.


\section{Comparison with observed AMDs}
 \label{subsec:gen_amd_comp}

In \citet{Adhikari2015}, the authors  demonstrated that the
observed AMD of Mrk 509 \citep{Detmers2011} can be explained by the absorber being under total constant pressure. The two dips obtained in the AMD model are in excellent agreement 
with the position of the observed AMD dips. 
Nevertheless, the normalization of the AMD in the  best fit model 
was higher by a  factor of  $\sim 30$ than the observed AMD normalization. 

\begin{figure}
\centering
\includegraphics[width=8.7cm]{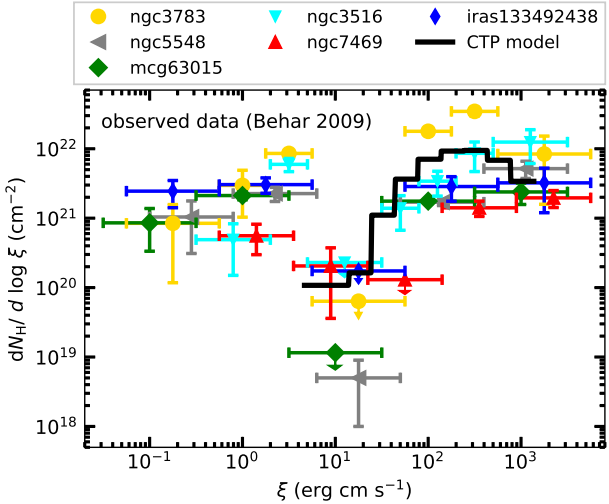}
\caption{The best fit AMD model (shown by a black histogram) 
computed with CTP assumptions in the  {\sc titan} code overplotted
with the observed AMD points obtained by \citet{Behar2009} for six 
Seyfert~1 galaxies. The AMD model parameters are: 
SED A with $\Gamma=2.0$,  $\xi_0=1.38 \times10^3$~erg~cm~s$^{-1}$, and $n_{\rm H,0}=10^{12}$ cm$^{-3}$, the latter two defined at the cloud surface.}
\label{fig:obs_amd_model}
\end{figure}

Fig.~\ref{fig:obs_amd_model} shows
a comparison between the observed AMDs for six Seyfert~1 galaxies
published by \citet{Behar2009} and our best fit model computed with SED A with $\Gamma=2.0$,  $\xi_0=1.38 \times10^3$~erg~cm~s$^{-1}$, and $n_{\rm H,0}=10^{12}$ cm$^{-3}$.
The model is binned in order to reconstruct the limited resolution of the observed AMD, which is 
always approximated by a step function \citep{Holczer2007}. The figure demonstrates that 
there is a very good agreement in the overall AMD 
normalizations ($\sim 4\times 10^{21}$~cm$^{-2}$) of those sources and our model.

Interestingly, observed multi-wavelength spectra obtained by different instruments are available for those six sources. The coverage is still poor, but the optical--X-ray index $\alpha_{\rm OX}$ was found for three sources by full spectral fitting with correcting for  absorption. These values  are respectively: 
--1.4 for NGC~3516, --1.31 for NGC~3783, --1.25 for NGC~5548 \citep{vasudevan2009}, and suggest the B type of SEDs (see Tab.~\ref{tab:param_gen_SEDs}). On the other hand, bolometric luminosities of those sources, 
determined from luminosity measured at $5100\AA$, span from 1 to $8 \times 10^{44}$~erg\,s$^{-1}$ (except NGC~3516, which has a luminosity $1.6 \times 10^{43}$~erg\,s$^{-1}$). Combining with  measured black hole masses, such luminosities suggest the accretion rates  relative to Eddington span  0.06 to 0.3, and this fact supports 
instead the  SED A case. Therefore, our result on AMD shape is not fully supported by the observed SED shape for those sources. We conclude here that there are some 
intermediate shapes of SEDs between A and B, which can provide the observed AMD shape. We plan to use them in our photoionization simulations in  future works.

\section{Discussion}
\label{sec:discussion}

Our computations indicate that for densities higher than $10^{10-11}$~cm$^{-3}$, free-free processes dominate over Compton processes, and both thermally unstable zones caused by ionization of heavy elements are almost completely suppressed. To reconstruct the full shape of AMD also in the  low ionization band as well, we should include radiation processes occurring in low-temperature dense gas.
Nevertheless, from our comparison in Fig.~\ref{fig:obs_amd_model}, the similarity in 
overall shape between the modelled and 
observed AMDs is very encouraging and clearly demonstrates that the CTP models 
are very successful in explaining the WA properties in Seyfert galaxies.
The requirement of high-density gas suppresses the high $\xi$ ($\sim 3000$) AMD dip which is not present in the data for  the \citet{Behar2009} sources. 

The observed dip is well modeled as a low-$\xi$ dip, but we see only its onset in the 
models. Nevertheless, we cannot increase the total column thickness since it is limited by the overall value of AMD normalization. 
In such a situation the {\sc titan} code does not allow us to reconstruct the AMD below $\xi~ \sim~ 4.5$ erg cm s$^{-1}$ (this value depends on initial cloud parameters). 
From the very beginning, {\sc titan} code was developed for  X-ray illuminated hot gas and 
the code is not suitable to reconstruct the gas below a temperature $8000$~K due to the 
lack of relevant radiation processes. The code does not take into account the condition that 
the  radiation is in thermal equilibrium the cold matter which contains dust. The same result was obtained by \cite{Goosmann2016} 
for NGC~3783's warm absorber. 
Therefore, only a part of the AMD drop is reconstructed here, but we show that both the position of the dip is well reproduced, and the overall normalization matches the observational data.

Prior,  AMD modeling by \citet{Stern2014} assumed that the WA 
is compressed by  radiation pressure (RPC) using the numerical code {\sc cloudy}.
However their RPC models did not reproduce the AMD dips which are caused by the thermal instability operating in the AGN photoionized outflow. In our
opinion, this is related to the issue that the transfer of radiation in {\sc cloudy}
is done using an escape probability method in one-stream approach, integrating gas from 
the front to back side of the cloud.  This method does not calculate the local radiation 
pressure while computing the cloud structure under CTP.
However in {\sc titan}, the radiative transfer is done using the ALI method,
where the proper source function term in the radiative transfer equation is taken into 
account. Since the computations are done on a much denser grid, 
and in the so called two-stream approach for radiative transfer computations,  
we are able to enter the few points in thermally unstable region which is demonstrated
in previous sections. 

It is known that the  precise description of the transition between the hot and cold thermally stable gas through the thermal 
instability zone is possible only 
when thermal conduction is taken into account \citep{Rozanska1999}, but in such a case 
radiative equilibrium becomes a second order differential equation, and computations 
become extremely complicated. Thermal instability occurs when the radiative
(energy) balance equation has three solutions: stable hot, stable cold and
intermediate/unstable. In a geometrically narrow zone inside a cloud, all three
solutions are available as a result of a subroutine solving local energy balance.
By default, the {\sc titan} code is using two-stream approximation, and it computes the
transfer from top and from the bottom and iterates to merge both results. But if we
want to follow only the hot or only cold solution, there is such option in the code.
Such simulations were done by \citet{Czerny2009}, but any further improvement can
be made when thermal conduction or evaporation will be included in the code
\citep{Rozanska1999}. Nevertheless, to combine the radiative transfer algorithm with 
thermal conduction and gas dynamics (i.e. evaporation) is very challenging subject,
and extremely time consuming, and such tasks are left to  future work
 with the development of 
new computational methods.
This method was also used by \citet{Goosmann2016} in case of Seyfert galaxy NGC~3783,
where the  authors were able to indicate the appearance of the thermal instability, but they did not reconstruct the depth of dips in the AMD. 
The authors found that an AMD model with ionization
parameter in the range 4000--8000 erg cm s$^{-1}$ represents
the observed AMD most closely, when the hot solution in {\sc titan} is adopted. 
Moreover, the authors do not discuss how their models 
are sensitive to the gas density adopted in the computations.

\section{Conclusions}
\label{sec:conclusion}

In this paper, we studied the  properties of WA clouds  in  AGN outflows  by 
performing the photoionization simulations using the numerical code 
{\sc titan} \citet{Dumont2000}. We considered two types of SEDs
illuminating  the warm absorber, which we divide into group A and group B \citep{adhikaripta2018}. Computations were done under an assumption that 
the total pressure inside the cloud i.e., $P_{\rm gas}+P_{\rm rad}$ is constant.
The temperature and ionization structures are computed to obtain the AMD for each model.

The temperature structure for different SED shapes indicates that
for models dominated by strong optical UV bump and weak X-ray component, i.e. SED A, the normalization  of the AMD derived with CTP models is of the order of a few~$\times 10^{21}$~cm$^{-2}$
 which is in excellent agreement with the observationally 
obtained average AMD normalization for six Seyfert~1 galaxies by \citet{Behar2009}. 

When the disk luminosity is equal to or smaller than that of the 
X-ray power law  i.e. SED B, the normalization of the AMD is 1.5 orders of 
magnitude higher than the normalization as determined  by observations,  and these SEDs are thus not favored for the modeled sources. We would like to point out, that while rejecting
SED B is physically motivated, the adoption of SED A does not exclude an intermediate spectral shape between A and B, which will also result in adequate AMD normalization. 
However, to extract the optimal spectral shape responsible for the 
observed AMD, we would need to  calculate a large number of models and to use machine learning method to compare of those models with observations. We plan to do this  in a future work.

We have shown, for the first time, how varying the SED shapes within group A  influences the AMD distribution due to having different 
shapes for the stability curve, a consequence of temperature structure, as presented in 
Sec.~\ref {sec:amd_dependence}. This is in agreement with previous studies on the influence 
of SED on the overall gas ionization stability curve due to illumination \citep{Chakravorty2009,Chakravorty2012,Dyda2017}.

For the first time we have specified  how the AMD shape depends on the 
intrinsic gas number  density. We conclude that the overall shape of the modeled  AMD agrees  with that from  observations
 for high density gas since a  high density is needed to suppress the dips in the AMD 
 at relatively high  values of the ionization parameter.
The agreement suggests  that the WA in Seyfert galaxies typically has a
density of the order of $10^{12}$~cm$^{-3}$ which is in agreement with the upper limit 
given by \citet{Elvis2017} for WA's  low-ionization phase.
The high-density requirement may indicate that the WA is formed from 
the upper layers of the disk atmosphere.
 Similar high-density gas was recently found to be present  in the 
broad and intermediate line regions \citep{hryniewicz2012,panda2018,adhikari2018}.
Nevertheless, the gas velocities inferred from broad lines in the optical/UV domain are considerably higher -- approximately  3 to 100 times higher 
than those from the WA medium as derived from individual observed lines. 
This fact may suggest different locations for both absorbers or different 
turbulent velocity gradients occurring in those media \citep{fukumura2010,fukumura2010L}.
The high density of the WA, as inferred from this paper, only means that the gas may originate from accretion 
disk atmospheres, which are dense. The question of what  speeds up the outflows, which 
we observe as a line-of-sight absorbing material,  is still not fully answered; it can be 
magnetic field or radiation pressure.

Our result for the high-density warm absorber puts tighter constraints on the wind location. 
Assuming a typical Seyfert~1  luminosity to be $10^{45}$~erg~s$^{-1}$, the location of 
such dense cloud could be estimated at $r \sim 10^{15}$~cm, which is about 
30 gravitational radii for a black hole mass of $10^8$ M$_{\odot}$. This location is very close to the SMBH and the question arises 
if such a warm absorber can survive in the very hostile innermost AGN region.
The derived distance of 30 gravitational radii based on the ionization parameter 
comes with the assumption of point source radiation. Using thin disc model 
it is straightforward to determine out that thermal hydrogen ionizing 
radiation is effectively emitted even four times farther compared to this 
radius. Thus in this region, the  radiation is extended, which makes distance 
estimation more uncertain.

X-ray magnetic flares above an accretion disk  were  proposed by several groups 
\citep{czernyflares2004,Ponti2004}, which may be a mechanism for 
formation of hot ionized gas clouds. Nevertheless, the estimation of the  lifetime for such 
clouds is not very optimistic.  As per  work by \citet{goosmann2006}, flares can 
live only 10 orbital periods, which gives  the lifetime of the order of several weeks (months). 
In such a situation the warm absorber will be highly variable and unstable,
in contradiction to what we really observe in 50\% of AGNs. On the other hand, the existence of clouds of that density and location were postulated by \citet{Lawrence2012} in order to explain the apparent the UV peak at 1100 \AA~ which is present in many AGN continua. More studies of the variability, including UV part where emission from these clouds is expected to be seen, may shed more light on the probability of this scenario.

\acknowledgments
{\small Special thanks  to Alex Markowitz for extensive editorial corrections
and to  Anne-Marie Dumont for many helpful discussions about the 
{\sc titan} code. This research was partially supported by Polish National Science 
Center grants No. 2015/17/B/ST9/03422, 2015/\-18/\-M/ST9/\-00541, 2015/\-17/\-B/ST9/\-03436,
and 2016/\-21/\-N/ST9/\-03311.}
\software{\sc titan\citep{Dumont2000}}
 
\bibliographystyle{apj}
\bibliography{refs}
\end{document}